\documentclass[twocolumn,prl,aps,superscriptaddress]{revtex4-1}
\usepackage[latin9]{inputenc}
\usepackage{mathtools}
\usepackage{amsbsy}
\usepackage{amssymb}
\usepackage{amstext}
\usepackage[unicode=true,pdfusetitle,
 bookmarks=false,
 breaklinks=false,pdfborder={0 0 1},backref=false,colorlinks=false]
 {hyperref}
\hypersetup{
 bookmarksnumbered=false,bookmarksopen=false}
\makeatletter
\def\maketitle{
\@author@finish
\title@column\titleblock@produce
\suppressfloats[t]}
\makeatother

\newcommand{\beq}{\begin{equation}}
\newcommand{\eeq}{\end{equation}}
\newcommand{\beqa}{\begin{eqnarray}}
\newcommand{\eeqa}{\end{eqnarray}}
\newcommand{\bra}{\langle}
\newcommand{\ket}{\rangle}

\usepackage{amsmath}
\usepackage{graphicx}
\usepackage{amssymb}
\usepackage{txfonts,color}
\makeatother

\begin{document}
\title{Tailored Ion Beam for Precise Color Center Creation}
\author{A. Tobalina}
\affiliation{Department of Physical Chemistry, University of the Basque Country UPV/EHU, Apartado 644, 48080 Bilbao, Spain}
\author{C. Munuera-Javaloy}
\affiliation{Department of Physical Chemistry, University of the Basque Country UPV/EHU, Apartado 644, 48080 Bilbao, Spain}
\author{E. Torrontegui}
\affiliation{Departamento de F\'isica, Universidad Carlos III de Madrid, Avda. de la Universidad 30, 28911 Legan\'es, Spain}
\affiliation{Instituto de F\'isica Fundamental IFF-CSIC, Calle Serrano 113b, 28006 Madrid, Spain}
\author{J. G. Muga}
\affiliation{Department of Physical Chemistry, University of the Basque Country UPV/EHU, Apartado 644, 48080 Bilbao, Spain}
\author{J. Casanova}
\affiliation{Department of Physical Chemistry, University of the Basque Country UPV/EHU, Apartado 644, 48080 Bilbao, Spain}
\affiliation{IKERBASQUE, Basque Foundation for Science, Plaza Euskadi 5, 48009 Bilbao, Spain}

\begin{abstract} 
We present a unitary quantum control scheme that produces a highly monochromatic ion beam from a Paul trap. Our protocol is implementable by supplying the segmented electrodes with voltages of the order of Volts, which mitigates the impact of fluctuating voltages in previous designs and leads to a low-dispersion beam of ions. Moreover, our proposal does not rely on sympathetically cooling the ions, which bypasses the need of loading  different  species in the trap --namely, the propelled ion and, e.g., a $^{40}$Ca$^+$ atom  able to exert sympathetic cooling-- incrementing the repetition rate of the launching procedure.  Our scheme is based on an invariant operator linear in position and momentum, which enables us to control the average extraction energy and the outgoing momentum spread. In addition, we propose a sequential operation to tailor the transversal properties of the beam before the ejection to minimize the impact spot and to increase the lateral resolution of the implantation.
\end{abstract}

\maketitle

\emph{Introduction.--} 
The creation of structural defects on crystalline structures is a relevant task on the road towards building solid-state quantum devices for groundbreaking applications such as nanoscale nuclear magnetic resonance (NMR)~\cite{Wu16, Degen17}. For instance, nitrogen-vacancy (NV) color centres in diamond~\cite{Doherty13} have attracted considerable attention in the last years for their potential to serve as a quantum platform for room temperature applications~\cite{Lovchinsky16}.  A possible fabrication technique of such structures consists on blasting a highly-pure diamond piece with nitrogen atoms such that, as they penetrate the diamond, they kick a carbon atom out of its location leading to a useful NV defect that behaves as a nanoscale magnetic compass~\cite{Smith19}. Creating shallow defects~\cite{Romach15} is specially convenient as it eases their control and enables relevant applications, such as using the NV as a nanoscale NMR probe of molecular samples in the diamond surface~\cite{Shi15, Aslam17, Kehayias17,Glenn18,Bucher20,Arunkumar21}. In this scenario, the depth at which the defect is created is directly related with the energy of the impacting atom, thus, achieving a flying atom with a narrow momentum distribution around a predetermined value would enable the creation of defects near the surface of the crystalline host. 

Many potential uses of those defects require high resolution placement, which can be achieved using masks \cite{Schukraft16} or pierced atomic force microscopes \cite{Pezzagna10,Riedrich-Moller15}. Creation of single defect centres with nanometer resolution has also been demonstrated using focused ion beams \cite{Schroder17}. A possible approach to produce an ion beam for high-precision color centre creation is to extract the ions from a linear Paul trap and direct them to the host though an electromagnetic lens \cite{Schnitzler09}. In these experiments the spread of the beam is significantly reduced by cooling down the ions. Unfortunately, nitrogen --along with many other interesting dopant species such as Germanium~\cite{Bhaskar17,Siyushev17}-- lacks the internal structure that allows for laser cooling~\cite{Jayich16}. Thus, dopants are typically cooled down through their interaction with different ion species that present a closed optical transition. In this manner, besides the dopant ion, this sympathetic cooling procedure \cite{Kielpinski00, Wubbena12} implies the ionization and loading of another ion such as $^{40}$Ca$^+$, necessarily decreasing the repetition rate of the implantation procedure. Moreover, the necessity of reaching an ion crystal for sympathetic cooling introduces additional issues that may add to the duration of the operation. For example, if more than two N$_2^{\,\,+}$ ions are trapped together with a $^{40}$Ca$^+$ the crystal will melt, in which case the operation has to be started from scratch~\cite{JacobThesis16}. Even when a two-ion crystal is formed, the heavier cooling ion that accelerates slower than the lighter dopant ion, may be located ahead in the shooting line. Consequently, an additional voltage sequence has to be applied to reorientate the crystal~\cite{Groot-Berning21}. 

Furthermore, once the dopant ion is loaded into the trap and cooled down below the Doppler limit, it is subjected to an electric field that propels it through the pierced endcap of the trap towards the crystalline target.  In the latest reported experiments~\cite{Jacob16,Groot-Berning19,Groot-Berning21}, the driving field is generated by applying a constant voltage of the order of kV to the endcap electrodes. Such large voltages generate a significant uncertainty in the extraction energy of the ion and lower the control over the depth of the newborn defect. 

In this Letter we propose a unitary quantum control scheme with a significant twofold benefit. On the one hand, our protocol decreases the ion-extraction-energy-uncertainty as it uses small voltages bearable by the segmented electrodes of a Paul trap. This leads to an ion beam with precise ejection velocity. On the other hand, our method requires no previous cooling of the dopant --thus, no cooling ions are needed-- leading to a significant increase on the ejection repetition rate. Our protocol relies on the existence of an invariant operator linear in momentum and position~\cite{Guasti03,Urzua19}, which is exploited here as a control tool to engineer the properties of the outgoing beam. 

\emph{Driving in the shooting direction.--} 
The axial motion of the trapped nitrogen ion is driven by a movable harmonic potential with controllable frequency. Therefore its dynamics along the coordinate $z$ is governed by
\beq
\label{hamz}
H_z(t)=\frac{p_z^2}{2m} + \frac 1 2 m \omega_z(t)^2 \left[z - z_0(t)\right]^2,
\eeq
where $m$ is the mass of the ion and the center of the trap $z_0(t)$ and its frequency $\omega_z(t)$ constitute the control parameters of the operation (the use of the subscript $z$ is motivated by a later treatment of the radial motion of the ion). We find that such evolution keeps constant the expectation value of the linear operator 
\beq
\label{inv}
I(t)= u(t) p_z - m \dot u(t) z + f(t),
\eeq
provided that the {\it auxiliary functions} $u(t)$ and $f(t)$ and the {\it control parameters} $\omega_z(t)$ and $z_0(t)$ obey the following equations,
\beqa
\label{auxeq1}
\ddot u(t) + \omega_z(t)^2 u(t) &=& 0,
 \\
 \label{auxeq2}
 \dot f(t) + m  \omega_z(t)^2 u(t) z_0(t) &=& 0.
\eeqa
The novel invariant presented here generalizes the operators used in previous works ~\cite{Muga20, Tobalina20} to introduce the displacement of the trap in the engineering of the driving.
The dynamical state of the ion can be conveniently written in terms of the eigenvectors of the invariant $I(t)$, while the corresponding evolution of $\bra z \ket$ and $\bra p_z \ket$ (also of $\Delta z$ and $\Delta p_z$) can be computed in terms of their initial values $\bra z \ket_0$ and $\bra p_z \ket_0$ (the explicit expressions are in the Supplemental Material~\cite{SM}). In fact, since our proposal does not rely on sympathetically cooling the ion near the absolute zero, we consider that the ion starts the operation at a thermal state characterised by $ \bra z \ket_0 =0 $, $ \bra p_z \ket_0 = 0$, $\bra zp_z +p_zz \ket_0 =0$, as well as 
\beqa
\label{statmoments}
 \bra z^2 \ket_0 = \frac{\hbar}{2 m \omega_{z,0}} \coth{\left(\frac{ \hbar \omega_{z,0}}{2 k_B T_0}\right)},\nonumber\\
 \bra p_z^2 \ket_0 = \frac{m \hbar \omega_{z,0}}{2} \coth{\left(\frac{\hbar \omega_{z,0}}{2 k_B T_0}\right)},
\eeqa
where $k_B$ is the Boltzmann constant and $T_0$ the initial temperature.

The invariance of $I(t)$ allows to establish useful relations between the expectation values of position and momentum operators at different times of the evolution. In particular, if $\dot u_{0,f} = 0$ (we use subscripts $f$ and $0$ for final and initial times), the motion of the ion yields
\beqa
\label{momscalingav}
\bra p_z \ket_f &=& R \bra p_z \ket_0 + P,\\
\label{momscalingdisp}
\Delta p_{z,f}^{\,\,2} &=& R\Delta p_{z,0}^{\,\,2} - 2 R P \bra p_{z,0} \ket,
\eeqa
where the coefficients $R=u_0/u_f$ and $P = (f_0-f_f)/u_f$ are tunable. Hence, by suitably fixing the boundary values of the auxiliary functions we can determine the average launching momentum and its outgoing spread, leading to the generation of a tailored ion beam. 

Along with $\dot u_{0,f} = 0$, a possible choice of boundary conditions (BC) leading to Eqs. \eqref{momscalingav} and \eqref{momscalingdisp} is 
\beqa
\label{bc1}
u_0=1, \hspace{.6cm} u_f = \frac 1 R, \hspace{.6cm}
f_0=0, \hspace{.6cm} f_f=- \frac P R.  
\eeqa
Further BC may determine other properties of the system. For example, using Eq.~\eqref{auxeq1}, we select the initial and final width of the trapping potential by imposing 
\beq 
\label{bc2}
\ddot u_0=- u_0 \omega_0^{\,\,2}, \hspace{.8cm} \ddot u _f = - u_f \omega_f^{\,\,2}, \hspace{.8cm} \dddot u_{0,f}=0.  
\eeq
Likewise, from Eq.~\eqref{auxeq2} the location $z_0(t_f)=d$ and the velocity  $\dot z_0(t_f)=v$ of the minimum of the potential at the end of the operation are specified through 
\beq 
\label{bc3}
\dot f_0=\ddot f_0=0, \hspace{.35cm} \dot f_f = - m u_f \omega_f ^{\,\,2} d,  \hspace{.35cm} \ddot f_f=- m u_f \omega_f ^{\,\,2} v.  
\eeq
Once $u(t)$ and $f(t)$ have been suitably designed, the control parameters leading to the tailored ion beam are found from Eqs. \eqref{auxeq1} and \eqref{auxeq2}. This inverse approach based on invariants is widely use in the context of shortcuts to adiabaticity \cite{Guery19}.

 Here we use polynomial auxiliary functions $u(t)$ and $f(t)$ (explicit expressions in~\cite{SM}) but note that any pair of auxiliary functions that satisfies the BC in Eqs. \eqref{bc1}, \eqref{bc2} and \eqref{bc3} yields the results predicted by Eqs. \eqref{momscalingav} and \eqref{momscalingdisp} and would lead to the target ion beam. Thus, our method provides infinite degrees of freedom to address additional goals or constraints. Here, we exploit this flexibility to reduce the maximum potential energy gained by the ion during its manipulation. In fact, the evolution of the ion is accurately described by  $H(t)$ provided that it never exceeds the depth of the trap. Note that the harmonic potential in Eq. \eqref{hamz} is a local approximation of the total potential generated within the Paul trap. Thus, reducing the energy peak avoids stages where the potential felt by the ion is not harmonic, thus ensuring the viability of the method. A detailed explanation on the design process can be found in the Supplemental Material~\cite{SM}. 
 
Avoiding transient departures from the harmonic regime poses a bound on the capacity of the method. Nevertheless, Fig.~\ref{f1} (a) demonstrates that our controls can be optimized to reduce transient energy peaks. In particular, in Fig.~\ref{f1} (a) we display some example energy peaks and show that, for trap depths achievable with current technology (around few eV \cite{Leibfried03, Schnitzler10}), we obtain controls that boost ions starting with temperatures as high as thousands of Kelvins to velocities of the order of km/s. Moreover,  the results shown in Fig.~\ref{f1} (a) admit further optimization tools that may lead to a beam with greater extraction energy for the same potential depth. 

\begin{figure}[t]
\begin{center}
\includegraphics[width=\linewidth]{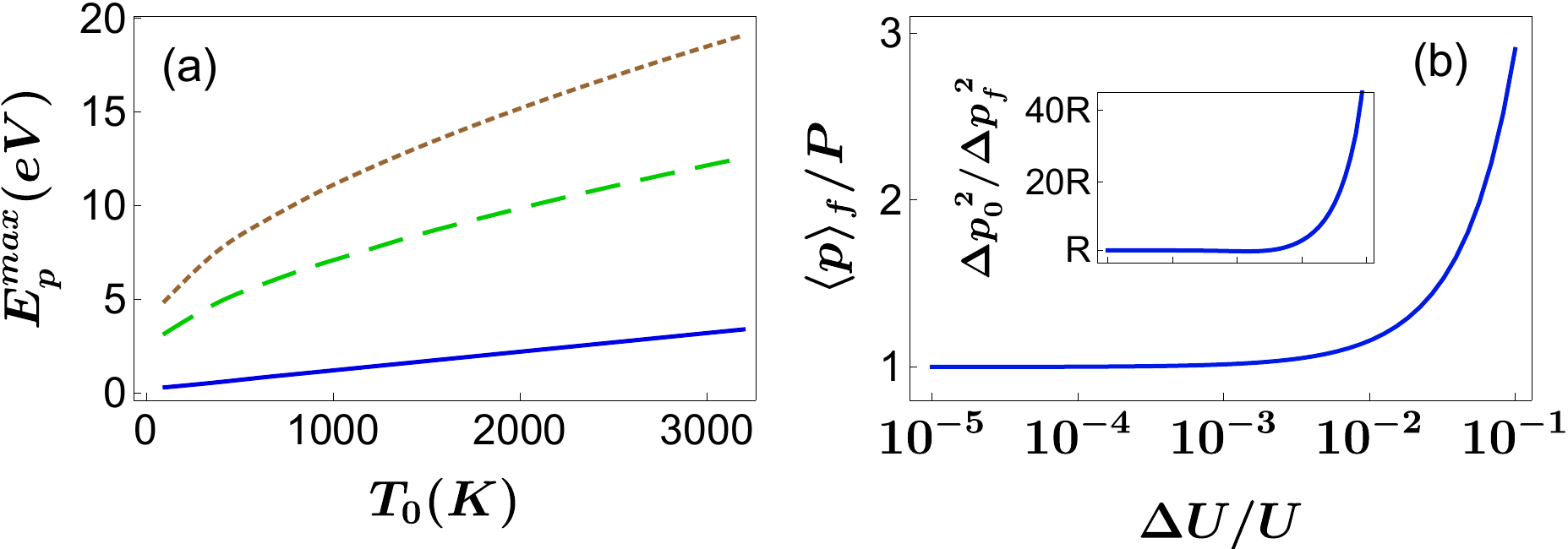}
\caption{\label{f1} The maximum value of the average potential energy $E_p=\frac 1 2 m \omega_z(t)^2 \left(z - z_0(t)\right)^2$  for different initial temperatures $T_0$ during a protocol that boosts the N$_2^{\,\,+}$ ion to launching velocities of $1$ km/s (solid blue), $3$ km/s (dashed green) and $5$ km/s (dotted brown) while reducing the width of the momentum spread by a factor of 2. (b) Average momentum and scaling parameter (inset) resulting from different relative uncertainties in the control voltages for an initial ion at $T_0=1000K$ and a target velocity of 5 km/s. All the operations are carried out in $t_f=0.94$ $\mu$s and the trap begins and ends with $\omega_0=\omega_f=(2\pi) \times 0.85$ MHz.}
\end{center}
\end{figure}

In previous reports on the creation of deterministic ion beams from Paul traps, imprecise voltages lead to fluctuating extraction energies that exceed by far the intrinsic uncertainty of the quantum system~\cite{Jacob16}. Our proposal admits an arbitrarily small scaling parameter $R$ --see Eqs.~(\ref{momscalingav}, \ref{momscalingdisp})-- so, formally, the momentum of the ejected ion may be as well defined as desired. However, uncertainties in the electrode voltages could spoil the outcome of the method presented here, so, next we evaluate the impact of imprecise electrodes in the outgoing beam. To this end, we consider a model~\cite{Furst14, Tobalina18} where the ion is driven by two adjacent electrodes that generate a potential $V(z,t)=e\left[\phi_1(z)U_1(t)+\phi_2(z)U_2(t)\right]$ with time-dependent voltages 
\beq
U_{1,2}(t)=\frac{ -  \phi'_{2,1} [z_0(t)] m \omega(t)^2}{( \phi''_2[z_0(t)]   \phi'_1[z_0(t)]  -  \phi'_2[z_0(t)]  \phi''_1[z_0(t)] )e},
\eeq
where $\phi_{1,2}(z)$ are dimensionless functions determined by the geometry of the electrodes and $e$ the fundamental electric charge. 

\begin{figure}[t]
\begin{center}
\includegraphics[width=\linewidth]{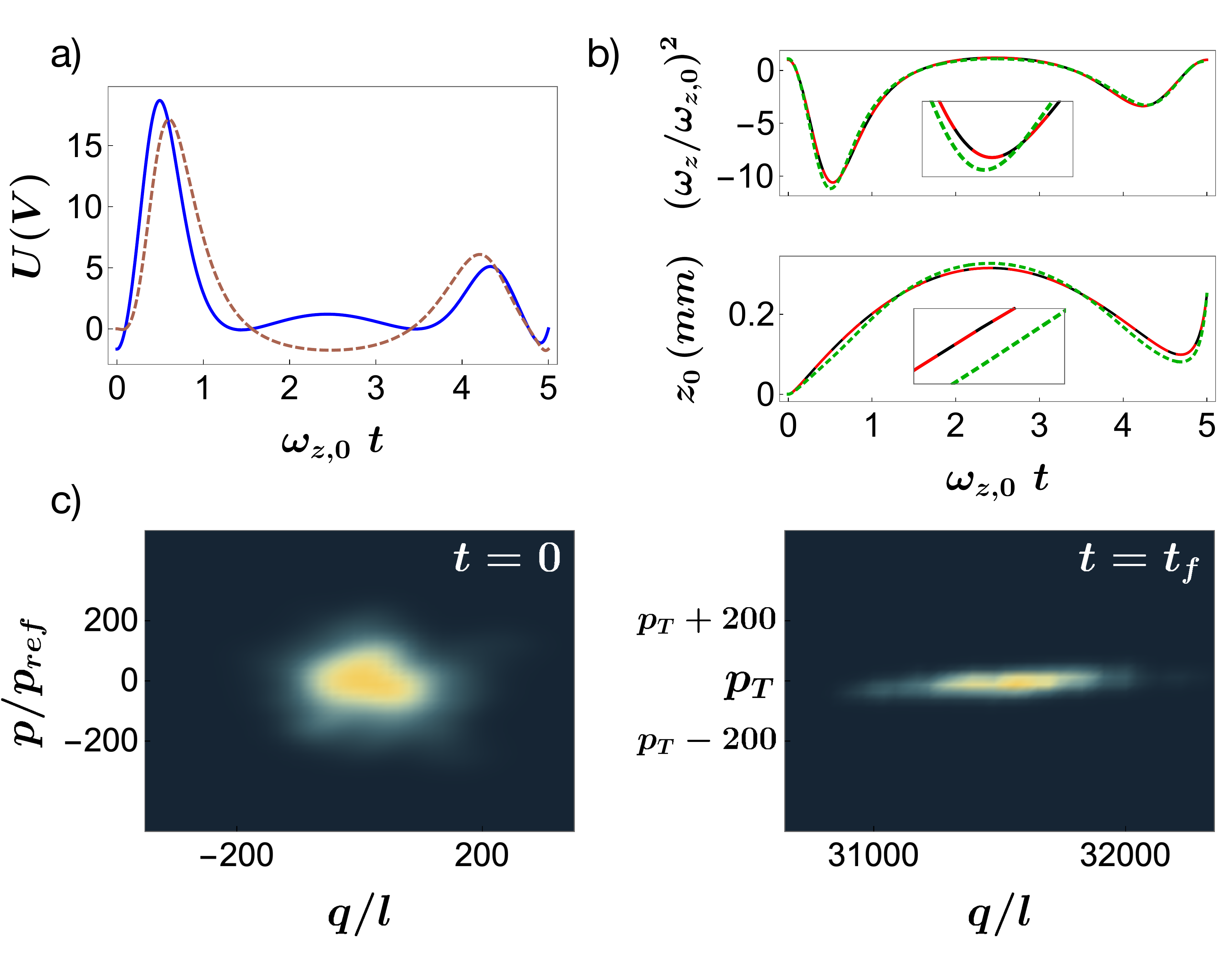}
\caption{\label{f2} a) Voltages to eject a single N$_2^{\,\,+}$ ion, $U_1(t)$ in solid blue and $U_2(t)$ in dashed brown, with average velocity of $5$ km/s and $R=1/5$, see Eq. \eqref{momscalingdisp}. The operation takes $0.94$ $\mu$s. The generated potential at boundary times has $\omega_0=\omega_f=( 2\pi) \times 0.85 $ MHz, ending at $d=250$ $\mu$m and with $v=10$ km/s. The geometric functions are modeled as Gaussians, $\phi_{1,2}(z)=0.2 \exp{\left[-(z-\beta_{1,2})^2 / (2 \sigma^2)\right]}$ with $\beta_1=0$, $\beta_2=250$ $\mu$m and $\sigma^2=200$ $\mu$m.  b) Control functions produced by the ideal voltages, solid black, and by voltages with relative uncertainty of $\Delta U/U=10^{-5}$, dashed red, and $\Delta U/U=10^{-1}$, dotted green. c) Initial and final Wigner functions for an ion starting with $T=1000$ K and driven by electrodes with a relative voltage uncertainty of $\Delta U/U=10^{-3}$. The reference length is $l=\sqrt{\hbar/(m \omega_{z,0})}$ and the reference momentum $p_{ref}=\hbar/l$.}
\end{center}
\end{figure}
Figure~\ref{f1} (b) displays the average extraction momentum and scaling parameter for simulations with different voltage uncertainties relative to the ideal scheme in Fig.~\ref{f2} (a), see Supplemental Material~\cite{SM} for a detailed explanation of the simulation. Figure~\ref{f2} (b) shows two examples of the control functions produced by inaccurate electrodes, together with the ideal driving, and Fig~\ref{f2} (c) displays the result of driving the ion with fluctuating electrodes through its Wigner function before (first panel) and after the manipulation (second panel). From Figure~\ref{f1} (b) one can learn that a relative uncertainty below $\Delta U / U =10^{-3}$ (we consider the same fluctuation in both electrodes, so $\Delta U / U \equiv \Delta U_{1,2} / U_{1,2}$) hardly alters the intended results. Note that in \cite{Schnitzler09} they reported a value of $\Delta U / U =10^{-5}$, while bigger errors would cause notable changes in the extraction velocity and the outgoing momentum spread. 

Another relevant feature of our protocol is its low voltage requirement. In particular, supplying the two electrodes with few Volts suffices to boost the ion to velocities of the order of km/s.
Notice also that the launching protocols are carried out in few trap oscillation periods, i.e., in a time scale of the order of $\mu$s. Furthermore, and compared with procedures that require to cool down the ion to near the absolute cero, our method provides a huge improvement in the repetition rate. More specifically, since our scheme does not need to trap and cool down a second/cooling ion, the bottleneck of the total operation is the loading of the molecular nitrogen which typically takes around 200 ms~\cite{JacobThesis16}. On the contrary, in standard methods, ion-crystal formation and arrangement issues limit the loading rate to 1 N$_2^{\,\,+}$ ion per minute when a  $^{40}$Ca$^+$  is already present in the trap~\cite{JacobThesis16}.

\emph{Transversal spread.--} A deterministic creation technique of single defects in diamond  controls not only the depth, but also the transversal location of the created defect. Here we tailor the potential acting on the radial directions to reduce the impact spot and thus increase the resolution of the implantation. We propose a sequential approach in which the transversal properties of the beam are addressed first, by suitably driving the radial motion of the ion, and then the launching operation designed in the previous section is carried out. 
The evolution of the ion in one of the radial degrees of freedom (both radial directions are symmetrically equivalent) is governed by
\beq
H_r(t)=\frac{p_r^2}{2m} + \frac 1 2 m \omega_r(t) r^2.
\eeq
In contrast to the longitudinal motion, the radial harmonic oscillators have a static center. In this case the design of the angular frequency $\omega(t)$ boils down to the case introduced in \cite{Muga20} where the control protocol stems from a single auxiliary function, Eq. \eqref{auxeq1}, as Eq. \eqref{auxeq2} is trivially satisfied for $f(t)=0$. The function $u(t)$ may be designed to tailor either the outgoing transversal position or momentum distributions. We consider two possibilities, narrowing down the final momentum spread and focusing the outgoing wave-packet, and compute the effect of each case in the the impact width of the beam. To determine the worth of the operation, we compare the results with the trivial evolution in which no potential manipulation takes place in the radial degrees of freedom. A detailed explanation of the design of the control function is given in~\cite{SM}. 

The spatial dispersion of the wave-packet during the free flight after some average time of flight $\bar t = m d_{ff} /  \bra p_z \ket$ (where $d_{ff}$ is the distance of the free flight and $m$ is the mass of the ion) reads
$\label{dispfreeflight}
\Delta r = \sqrt{\bra r^2 \ket_0+ \bar t\,^2 \bra p_r^{\,\,2} \ket_0/m}
$
~\cite{SM}.
As explained before, the initial state of the ion is thermally distributed and its statistical moments have the form written in Eq.~\eqref{statmoments} but with the corresponding frequency $\omega_{r,0}$.
Intuitively, focusing the wave-packet may seem the right choice, inasmuch as the goal is to reduce the spatial width at the impact point. However, it turns out that the scattered radial momenta blow the state up during the free flight, actually deteriorating the result obtained from the trivial evolution. On the other hand, appropriately scaling down the radial momenta, even at the expense of starting with a wider state, improves considerably the impact spot one would get form the initial thermal state as it is shown by Fig.~\ref{f3} (a).
\begin{figure}[t]
\begin{center}
\includegraphics[width=\linewidth]{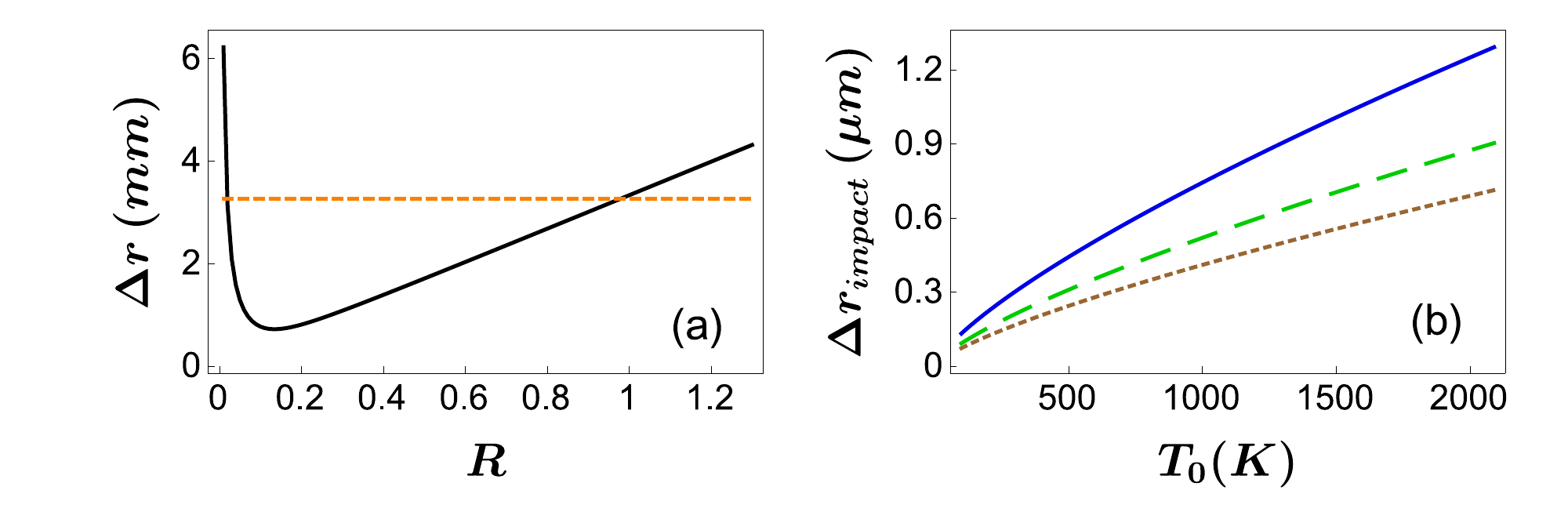}
\caption{\label{f3} (a) Radial dispersion after a free flight distance of $d_{ff}=300$ mm for invariant based drivings with different momentum scaling parameters, see Eq. \eqref{momscalingdisp}. Benchmark dispersion for a thermal state at $T=1000$ K (dashed orange), which is also the starting point for the invariant based protocol. (b) Dispersion at the impact spot for ions that have gone through an einzel lens with focal length $F=13$mm with velocities of 5 km/s (solid blue), 10km/s (dashes green) and 50 km/s(dotted brown).  Other parameters are $\omega_r = (2\pi) \times 1.4 $ MHz and $t_f=\omega_r^{\,\,-1}$, $t_f=0.94$ $\mu$s and $\bra p_z \ket = 5 \times 10^4$ m/s.}
\end{center}
\end{figure}

Additionally, if we consider that the ion goes through an electromagnetic lens before the impact, which is a typical setting in state-of-the-art experiments, the cross area of the beam may be reduced even further. Assuming that the target is placed at the focal point of an einzel lens \cite{Schnitzler10}, where the outcome has minimum width, and provided that the position and momentum of the incoming beam are Gauss distributed (which holds for the thermal state) the impact spread can be computed as~\cite{JacobThesis16}
\beq
\Delta r_{impact}=\frac{F \Delta r \Delta \alpha}{\sqrt{\left(F-d_{ff}\right)^2 \Delta \alpha^2 + \Delta r^2}},
\eeq
where $F$ is the focal length, the free flight distance $d_{ff}$ is the separation between the launching point and the principal plane of the lens and the beam divergence $\Delta \alpha$ can be computed from the ingoing radial width as $\Delta \alpha = \Delta r / d_{ff}$. Figure~\ref{f3} (b) displays the impact spot radius in terms for various initial temperatures and launching velocities. It exposes the bigger advantage of cooling the ion to sub-Doopler temperatures, which is the capacity to reach nanoscale lateral resolution \cite{Groot-Berning19,Groot-Berning21}. In our case, the radial driving allows to reduce the spot below one $\mu$m for reasonable initial temperatures. This result, however, could be further improved without invoking the time consuming sympathetic cooling, for example  through voltage sequences devoted to reduce the energy of the ion during the loading process, via more elaborate focusing setups (for example concatenating more lenses) or exploiting the flexibility of our method to find optimal trajectories~\cite{Stefanatos10,Chen11}.

\emph{Conclusions.--} We have developed a theoretical proposal to produce, from a Paul trap, a highly monochromatic beam of N$_2^{\,\,+}$ ions that could be accurately implanted into a diamond host to generate shallow NV color centers. Even for drivings produced by realistic/fluctuating electrodes, our invariant based protocol limits the uncertainty of the extraction energy to the value determined by the thermal distribution and even reduces it. It also provides a substantial improvement on the repetition rate over implantation processes that rely on cooling down the ions, so it would facilitate the deterministic production of large amounts of defects and its expansion to industrial use. A sequential use of our method can also tailor the radial properties of the outgoing beam to lower its dispersion.

\begin{acknowledgements}
\textit{Acknowledgements.--}  
Authors acknowledge insightful discussions with Karin Groot-Berning. We acknowledge financial support from Spanish Government via PGC2018-095113-B-I00 (MCIU/AEI/FEDER, UE), PGC2018-101355-B-100 (MCIU/AEI/FEDER, UE) and EUR2020-112117, from Basque Government via IT986-16, as well as from QMiCS (820505) and OpenSuperQ (820363) of the EU Flagship on Quantum Technologies, as well as from the EU FET Open Grant Quromorphic (828826). C.M.-J. acknowledges the predoctoral MICINN grant PRE2019-088519. J.C. acknowledges the Ram\'on y Cajal program (RYC2018-025197-I) and support from the UPV/EHU through the grant EHUrOPE. E.T. acknowledges financial support from the Spanish Government through PGC2018-094792-B-I00 (MCIU/AEI/FEDER,UE), CSIC Research Platform PTI-001, CAM/FEDER Project No. S2018/TCS-4342 (QUITEMAD-CM), and  by Comunidad de Madrid-EPUC3M14. 
\end{acknowledgements}

\pagebreak
\title{Tailored ion beam for deterministic color center creation.\\ SUPPLEMENTAL MATERIAL}

\maketitle

\section{Dynamical state of the ion and its statistical moments}
The state of the N$_2^{\,\,+}$ ion during the launching operation may be written in terms of the eigenvectors of the invariant, $I(t) | \phi _{p_{z,0}} (t) \ket =u_0  p_{z,0} | \phi _{p_{z,0}} (t) \ket$, with corresponding wave-functions 
\beqa
\label{eigenfuncinv}
\phi_{p_{z,0}} (z,t) &=&  \bra z | \phi_{p_{z,0}}(t) \ket \nonumber\\
&=& \frac{e^{i \varphi_{p_{z,0}}(t)}}{\sqrt{\hbar}} e^{i \left[ (u_0 p_{z,0} - f) q + m \dot u  z^2/2\right]/\hbar u}.
\eeqa
The phase 
\beq
e^{ i \varphi _{p_{z,0}}(t)} = \sqrt{\frac{u_0}{u}} e ^{ -\frac i {2 \hbar m} (\mathcal{I}_1 p_{z,0}^2 - 2 \mathcal{I}_2 p_{z,0} - \mathcal{I}_3)}
\eeq
given here in terms of the integrals $\mathcal I_1= \int dt/u^2$, $\mathcal I_2 = \int dt f /u^2$ and $\mathcal I_3 = \int dt (\dot f / \ddot u - f^2 /u)/u $, is chosen so that the function in Eq. \eqref{eigenfuncinv} satisfies $i \hbar \partial_t \phi _{p_{z,0}}(z,t) = H(t) \phi _{p_{z,0}}(z,t)$, i.e., it becomes a solution of the time-dependent Schr\"odinger equation. Thus, the state of our system at any given time can be expressed as 
\beqa
\label{wavefunc}
\psi(z,t)& = &\int dp_{z,0} \left( \frac{u_0}{\hbar u}\right)^{\frac 1 2} \exp\bigg[\frac i {\hbar u} \bigg( \frac{m \dot u} 2 z^2  + (u_0p_{z,0} - f) z\nonumber\\
& - &\frac u{2m} (\mathcal{I}_1 p_{z,0}^2 - 2 \mathcal{I}_2 p_{z,0} - \mathcal{I}_3)\bigg)\bigg] \bra p_{z,0}|\psi (0)\ket,
\eeqa
where the initial momentum $p_{z,0}$ has been used as the integration variable to expand the wave-function.
The corresponding expectation values for the position and momentum operators, given in terms of their initial values, read
\beqa
\label{q1}
\bra z \ket  &=& \int dz \psi^*(z,t) z \psi (z,t)\nonumber \\
&=& \frac u {u_0} \bra z \ket_0 + \frac {u_0 u} m \mathcal I_1 \bra p_z \ket_0 - \frac u m \mathcal I_2, 
\eeqa
\beqa
\label{p1}
\bra p_z \ket  &=& \int dz \psi^*(z,t) (- i \hbar \frac {\partial}{\partial z}) \psi (z,t) = \frac{m \dot u}{u_0} \bra z \ket _0\nonumber \\
 &+&  \left( \frac {u_0} u + u_0 \dot u \mathcal I_1 \right) \bra p_z \ket_0 - \dot u \mathcal I_2 + \frac{(f_0-f_f)}{u} .
\eeqa
Notice that the solution for $\bra z \ket$ is, according to Ehrenfest's theorem, a solution for the classical equation of motion
\beq
\label{claseqmot}
\ddot z + \omega_z(t)^2 z = \omega_z(t)^2 z_0(t),
\eeq
with control parameters that obey Eqs. (3) and (4) of the main text.
Thus, by computing the expectation value of the position for a state described by \eqref{wavefunc} we have found an analytical solution (in terms of some integral forms) for the time-dependent forced harmonic oscillator:
\beq
z (t)  =  \frac {u(t)} {u_0} z_0 + \frac {u_0 u(t)} m \mathcal I_1 p _{z,0} - \frac {u(t)} m \mathcal I_2.
\eeq

From here the second order moments may be computed using the Liouville theorem, which states that the phase-space distribution function is constant along the trajectories of the system. The theorem regards classical dynamics, however, the Wigner  function $W(z,t)$, i.e., the quasiprobability distribution that describes our quadratic system in phase-space, evolves classically and thus fulfills Liouville's theorem, leading to
$
W_0(z_0,p_{z,0}) = W_t (z,p_z),
$ 
allowing us to easily obtain higher order moments for $q$ and $p$ from their expectation values:
\beqa
\label{q2}
\bra z^2 \ket &=& \iint dz dp_z W_t(z,p_z) z^2 = \left(\frac u m \mathcal I_2\right)^2 +  \left(\frac{u}{u_0} \right)^2\bra z^2 \ket_0 \nonumber\\
&-& 2 \frac {u^2}{m u_0} \mathcal I_2 \bra z \ket_0 - 2  \frac {u_0 u} m  \mathcal I_1 \mathcal I_2 \bra p_z \ket_0 + \left( \frac{u_0 u}{m} \mathcal I_1\right)^2 \bra p_z^2 \ket_0   \nonumber\\ 
&+&   \frac{u^2} m \mathcal I_1 \bra zp_z + p_zz\ket_0,
\eeqa
\beqa
\label{p_z^2}
\bra p_z^2 \ket & = & \iint dz dp_z W_t(z,p_z) p_z^2 = \dot u ^2 \mathcal I_2 ^2 + 2 \frac {\dot u} u \mathcal I_2 \Delta f + \left(\frac{\Delta f}{u} \right)^2 \nonumber\\  
& - &  \left( 2m \frac{\dot u^2}{u_0} \mathcal I_2 + 2m \frac{\dot u}{u_0 u} \Delta f \right)\bra z \ket_0 + \left(\frac{m \dot u}{u_0} \right)^2 \bra z^2 \ket_0  \nonumber\\
&-& \left( 2 \dot u^2 u_0 \mathcal I_1 \mathcal I_2 + 2 \frac{\dot u}{u} \mathcal I_2 + 2 \frac{u_0 }{u^2} \Delta f + 2 \frac{u_0 \dot u}{u}  \Delta f \mathcal I_1 \right) \bra p_z \ket_0\nonumber\\ \nonumber
&+&\left( u_0 ^2 \dot u^2 \mathcal I_1^2 + \frac{u_0^2}{u^2} + 2 \frac{u_0^2 \dot u}{u} \mathcal I_1 \right) \bra p_z^2 \ket_0 \\
&+&  \left( m \dot u^2 \mathcal I_1 + \frac{m \dot u} u \right) \bra zp_z + p_zz\ket_0.\\ \nonumber
\eeqa
\section{Radial dispersion during the free flight}
Liouville's  theorem also allows us to easily compute the result of the free motion of the flying atom, as no force is acting on the quantum system. 
The average radial position mimics classical free motion,
\beq
\bra r \ket = \bra r \ket_0 + \frac{\bra p_r \ket_0}{m} t,
\eeq
and its dispersion reads 
\beqa
\bra r^2 \ket &=& \iint dr dp_r W_t(r,p_r) r^2 \nonumber\\
&=& \bra r^2 \ket_0 + \frac{\bra p_r^2 \ket_0}{m} t + \frac{\bra rp_r + p_r r \ket_0}{m}t.
\eeqa
From here, and considering the statistical moments of the initial thermal state, the radial dispersion of the flying ion is
\beq
\Delta r^2=\bra r^2\ket - \bra r \ket ^2 = \bra r^2 \ket =  \bra r^2 \ket_0 + \frac{\bra p_r^2 \ket_0}{m} t.
\eeq
\section{Invariant based design of the control parameters}
Here we take an inverse engineering approach to deduce the driving that produces the desired ion beam. In general, such design method requires three ingredients: (i) the operator that describes the evolution of the system, i.e., the Hamiltonian, (ii) a property of the system that remains constant under that evolution, i.e., the expectation value of the invariant operator, and (iii) some connection between the time-dependent functions that define both of them, which in our case are the equations that relate the control parameters $\omega_z(t)$ and $z_0(t)$ with the auxiliary parameters $u(t)$ and $f(t)$, Eqs. (3) and (4) in the main text. 

Once we have identified the necessary elements, we must translate the desired outcome into boundary conditions (BC) for the auxiliary functions. The BC leading to the tailored ion beam are given in Eqs. (8), (9) and (10) of the main text. Finally, all that remains is to propose functions for the auxiliary parameters that satisfy the deduced BC and compute the control parameters from Eqs. (3) and (4), which read
\beq
\omega_z(t)= \sqrt{-\frac{\ddot u(t)}{u(t)^2}},  \hspace{.6cm} z_0(t)=\frac{-\dot f(t)}{m u(t) \omega_z(t)^2}.
\eeq
Notice that, given these definitions, we might encounter discontinuities that render unfeasible control parameters. In particular, both $\omega_z(t)$ and $z_0(t)$ diverge whenever $u(t)=0$, and, additionally, $z_0(t)$ diverges if $\omega_z(t)$ becomes cero. To avoid discontinuities caused by the zeros of $u(t)$, we will design a function that is positive at all times. On the other hand, the zeros of $\omega_z(t)$ are harder to avoid, so instead of seeking a design of the frequency that never changes sign, we will force $\dot f(t)$ to match the zeros of $\omega_z(t)$, so that the resulting trajectory $z_0(t)$ is free of divergences. 

We use polynomial functions for the auxiliary parameters, 
\beq
u(t)=\sum_{j=0}^{10} a_j (t/t_f)^j, \hspace{0.8cm} f(t)=\sum_{k=0}^{11} b_k (t/t_f)^k.
\eeq
The coefficients will be fixed in three different steps. First  $a_{0-7}$ and $b_{0-5}$ are determined by the BC that the functions must meet. Then $a_8$ is used to maintain $u(t)$ positive at all times (we force $u(t_f/2)$ to have a certain value that hinders the change), while $b_{6-9}$ are used to impose the necessary zeros in $\dot f(t)$ to avoid divergences in $z_0(t)$. Finally, $a_{9,10}$ and $b_{10,11}$, two coefficients in each auxiliary function, are used to minimise the maximum value of the average potential energy of the ion using the 'fminseach' tool of Matlab.

\vspace{.5cm}
\section{Simulation of the dynamics}

The simulation of the evolution of the quantum system is considerably simplified (regarding computational effort) by the fact that the Wigner function in a quadratic potential evolves classically.

Thus, to simulate the dynamics of the ion we define a large amount of points, distributed in phase-space according to the Wigner function, which for the initial thermal state of the harmonic oscillator reads 
\beq
W(z,p_z)=\frac{1}{2 \pi \sqrt{\bra z^2 \ket_0 \bra p_z^2 \ket_0}}\exp{\left[\frac 1 2\left(  \frac {z^2}{ \bra z^2 \ket_0} + \frac{p_z^2}{\bra p_z^2\ket_0}\right)\right]},
\eeq
and evolve then individually using the classical equation of motion in Eq. \eqref{claseqmot}.

\end{document}